# Spatial Domain Watermarking Scheme for Colored Images Based on Log-average Luminance

Jamal A. Hussein

**Abstract**—In this paper a new watermarking scheme is presented based on log-average luminance. A colored-image is divided into blocks after converting the RGB colored image to $YC_bC_r$ color space. A monochrome image of 1024 bytes is used as the watermark. To embed the watermark, 16 blocks of size 8X8 are selected and used to embed the watermark image into the original image. The selected blocks are chosen spirally (beginning form the center of the image) among the blocks that have log-average luminance higher than or equal the log-average luminance of the entire image. Each byte of the monochrome watermark is added by updating a luminance value of a pixel of the image. If the byte of the watermark image represented white color (255) a value α is added to the image pixel luminance value, if it is black (0) the α is subtracted from the luminance value. To extract the watermark, the selected blocks are chosen as the above, if the difference between the luminance value of the watermarked image pixel and the original image pixel is greater than 0, the watermark pixel is supposed to be white, otherwise it supposed to be black. Experimental results show that the proposed scheme is efficient against changing the watermarked image to grayscale, image cropping, and JPEG compression.

**Index Terms**—Watermarking, Luminance, Gray-scale, log-average luminance, spatial domain

──────────  ◆  ──────────

## 1. INTRODUCTION

Watermarking is the methods of embedding information like personal data or logo into digital contents like images, audio or video. It can be detected or extracted later to prove the correctness of the embedded information in the digital media. The watermarking is essentially used in copyright protection, ownership proving, and integrity verification.

The watermark can be embedded into the image in spatial domain or domain frequency. In spatial domain techniques the watermark is embedded directly into the pixel data [1], [2], and [3]. In frequency domain techniques the image data is first converted to frequency domain using transforms such as DCT, DFT, or DWT. The watermark is embedded to the frequency domain coefficients and then the inverse transform is performed to restore the watermarked image.

In this paper a new scheme is proposed to embed a 32X32 monochrome watermark image (1024 bytes) in spatial domain based on log-average luminance. First, the original image is converted from RGB to $YC_bC_r$ color space and then the image is divided into 8X8 blocks. The log-average luminance is computed for the image and for each block of the image. Blocks that are used in watermark embedding are those blocks that have a log-average luminance equal or greater than the log-average luminance of the entire image. Since only 16 blocks are needed to embed the watermark, we choose them spirally among the selected blocks beginning from the center of the images. The monochrome watermark image only has two colors, black and white. Each byte of the watermark image is embedded into the original image by changing the pixel's Y value of the selected blocks. The α is added to Y value of the pixel if the watermark pixel is white (255) or α is subtracted if the color is black (0), where α is an integer value.

To extract the watermark the same steps of embedding watermark is performed except of the embedding the extracting operation is performed by subtracting the luminance value of the watermarked image pixel from the luminance of the original image pixel, if the result is greater than or equal zero the watermark pixel is supposed to be white, otherwise, when the result of subtraction is less than zero, the watermark pixel supposed to be block, see figure 1.

────────────────

- *J.A.Hussein is with the Computer Department, College of Science, University of Sulaimani. Sulaimani, Kurdistan-Iraq.*

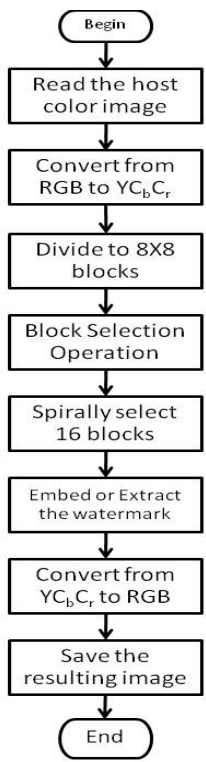

Figure 1. Watermark embedding and extracting processes

## 2. PROPOSED SCHEME FOR WATERMARKING

### 2.1 YC$_b$C$_r$ Conversion

In YC$_b$C$_r$, the Y component represents the luminance (brightness), while C$_b$ and C$_r$ represent blue-difference and red-difference respectively.

Since the watermark is added to the luminance, the RGB color space of the image should be converted to YC$_b$C$_r$ color space. The Y component is used later to embed the watermark.

Y = 0.299 * R + 0.587 * G + 0.114 * B
C$_b$ = 0.596 * R - 0.275 * G - 0.321 * B
C$_r$ = 0.212 * R - 0.523 * G - 0.311 * B

To convert YCbCr to RGB

R = Y + 0.956 * C$_b$ + 0.620 * C$_r$
G = Y - 0.272 * C$_b$ - 0. 647 * C$_r$
B = Y - 1.108 * C$_b$ + 1.705 * C$_r$

### 2.2 Log-average Luminance

Since the Y component of the YCbCr color space represents the luminance of the pixel. We can find the log-average luminance of the entire image or block by using the following equation:

$$L_{avg} = \exp(\sum \log(\delta + Y_{x,y})/N)$$

Where

$L_{avg}$: Log-average luminance

$Y_{x,y}$: Luminance of the pixel at x,y

$\delta$: A small value to avoid taking the log of a completely black pixel whose luminance is zero

N: The number of pixel in the image

### 2.3 Block Selection Operation

The image is divided into 8X8 blocks. The log-average luminance is computed for the entire image and for each block. The candidate blocks used in watermark embedding are those blocks that have log-average luminance greater or equal than the log-average luminance of the entire image. To embed the 1024 pixel monochrome watermark image, 16 blocks are needed. These blocks are chosen beginning from the center of the image in a spiral manner as shown in figure 2.

### 2.4 Watermark embedding

The monochrome watermark image only consists of black (0) and white (255) pixels. So to add the watermark, if the pixel is white the α is added to the pixel luminance of the original image, for black pixels the value α is subtracted see figure 3.

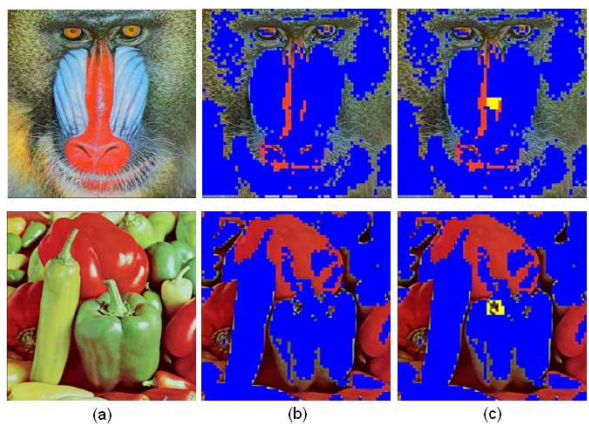

Figure 2.
(a) the image (b) selected blocks (blue) (c) watermarked blocks (yellow)

$$Y'_{x,y} = \begin{cases} Y_{x,y}+\alpha & \text{where } W_{x,y}=255 \\ Y_{x,y}-\alpha & \text{otherwise } (W_{x,y}=0) \end{cases}$$

$Y'_{x,y}$: Pixel luminance value after addition

$Y_{x,y}$: Original pixel luminance value

α: addition factor

$W_{x,y}$: watermark pixel value (0 or 255)

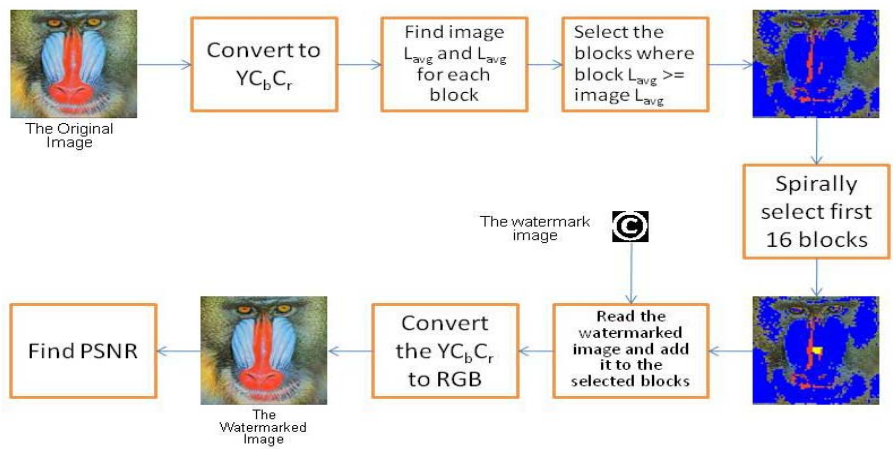

Figure 3. the embedding process

The Peak Signal to Noise Ratio PSNR is used as a quality measurement of the imperceptibility of the watermarked image when compared with the original image. The higher the PSNR value the best the watermarked image visual quality.

$$\text{PSNR(dB)} = 10\log_{10}(m^2 \frac{\max(I(x,y))^2}{\sum(I(x,y)-I_w(x,y))^2})$$

## 2.5 Watermark Extracting

The watermark is extracted in a non-blind approach, i.e both original image and watermarked image are required in order to exctract the watermark. to produce the watermarks' pixels, we first convert each of original and watermarked images to $YC_bC_r$ color space, the blocks selection operation is performed as in section 2.1 and each Y component of the watermarked pixel is subtracted from the corresponding Y value of the original pixel. If the result is positive or zero the extracted watermark pixel is considered to be white, otherwise black pixel is considered. This operation is continued until all watermarked pixels are produced see figure 4.

$$W'_{x,y} = \begin{cases} 255 & \text{where } Y^w_{i,j} - Y_{i,j} \geq 0 \\ 0 & \text{otherwise } (Y^w_{i,j} - Y_{i,j} < 0) \end{cases}$$

Where

$W'_{x,y}$ : The extracted watermark xy pixel

$Y^w_{i,j}$ : The ij pixel Y value of watermarked image

$Y_{i,j}$ : The ij pixel Y value of original image

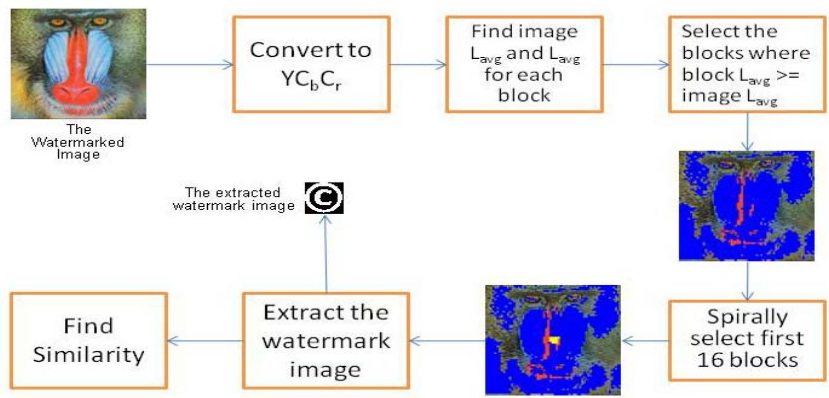

Figure 4. the extracting process

## 3. Quality Measurement

Similarity measure is used to compare the extracted watermark W' to the originally embedded watermark W.

$$\sigma = \sum_x \sum_y S_{x,y}$$

$$\text{where} \quad S_{x,y} = \begin{cases} 1 & \text{if } W'_{x,y} = W_{x,y} \\ 0 & \text{if } W'_{x,y} \neq W_{x,y} \end{cases}$$

The similarity $\sigma$ varies in the interval [0, 1]; the value in the interval (0.5, 1.0] shows that the extracted watermark W' is matching the embedded watermark W, this means that the original image has been watermarked with W.

## 4. Experimental Results

The results are produced from tests applied on different color images ('baboon', 'peppers', and 'lena'). The block size is taken as constant (8x8).
Table 1 show the quality measure PSNR and the Similarity for different images using different tests such as Cropping, JPEG compression, and change the colored image to grayscale image.
As we see, the similarity is not affected by image cropping and color image to grayscale conversion. In compression, the quality of the watermark is decreased as the compression factor is increased.

Table 1. Various tests on different colored images where α = 3

| Tests | Images | | | | | |
|---|---|---|---|---|---|---|
| | Baboon | | Peppers | | Lena | |
| | PSNR | $\sigma$ | PSNR | $\sigma$ | PSNR | $\sigma$ |
| No change | 62.499 | 1.0 | 62.499 | 1.0 | 62.499 | 1.0 |
| Cropping | 56.478 | 1.0 | 56.478 | 1.0 | 56.478 | 1.0 |
| Compression (CF = 0.75) | 26.035 | 0.676 | 30.112 | 0.670 | 32.512 | 0.689 |
| To grayscale | - | 1.0 | - | 1.0 | - | 1.0 |

## 5. Conclusion and Future works

In this paper a color image watermarking technique is proposed based on log-average luminance. This technique is tested on various images. The watermark is a monochrome image. The best blocks are selected spirally, beginning from the center, among those blocks that have log-average luminance greater than or equal to the log-average luminance of the entire image.
Experimental test results show best watermark invisibility when the addition factor α is 3. Many tests are performed on different color images and show some robustness against various attacks.
More robustness can be achieved by adding the watermark in frequency domain using transforms like (DFT, DCT, or DWT). We also can use grayscale or color watermark instead of the monochrome watermark used in this paper.

**Jamal A. Hussein** MSc in Computer Science 2007. Member of Scientific Committee in Computer Department at College of Science / University of Sulaimani. He has some research papers in the Watermarking field.